\documentclass[rnote]{aa}
\usepackage{color}
\usepackage{graphicx}
\usepackage{lscape}
\usepackage{natbib}
\usepackage[scaled]{helvet}
\usepackage[varg]{txfonts}
\usepackage{url}
\usepackage{xspace}
\bibpunct{(}{)}{;}{a}{}{,}
\newcounter{Rco}

\newcommand{\Jonw}[3]{\mbox{\ion{#1}{#2}~$\lambda\,#3$\,\AA}\xspace}
\newcommand{\Jonww}[3]{\mbox{\ion{#1}{#2}~$\lambda\lambda\,#3$\,\AA}\xspace}

\newcommand{\loggw}[1]{\mbox{$\log g\hspace{-0.5mm} =\hspace{-0.5mm}  #1$}}

\newcommand{\Teff}{\mbox{$T_\mathrm{eff}$}\xspace}
\newcommand{\Teffw}[1]{\mbox{$\Teff\hspace{-0.5mm} =\hspace{-0.5mm} #1 \,\mathrm{K}$}}

\newcommand{\re}{\object{RE\,0503$-$289}\xspace}
\begin{document}

\title{Stellar laboratories}
\subtitle{V. The \ion{Xe}{vi} ultraviolet spectrum and the xenon abundance \\
             in the hot DO-type white dwarf \re
             \thanks
             {Based on observations made with the NASA-CNES-CSA Far Ultraviolet Spectroscopic Explorer.
             }$^,$
             \thanks
             {Based on observations with the NASA/ESA Hubble Space Telescope, obtained at the Space Telescope Science 
             Institute, which is operated by the Association of Universities for Research in Astronomy, Inc., under 
             NASA contract NAS5-26666.
             }
         }
\titlerunning{Stellar laboratories: the \ion{Xe}{vi} spectrum}

\author{T\@. Rauch\inst{1}
        \and
        D\@. Hoyer\inst{1}
        \and
        P\@. Quinet\inst{2,3}
        \and
        M\@. Gallardo\inst{4}
        \and
        M\@. Raineri\inst{4}
        }

\institute{Institute for Astronomy and Astrophysics,
           Kepler Center for Astro and Particle Physics,
           Eberhard Karls University,
           Sand 1,
           72076 T\"ubingen,
           Germany \\
           \email{rauch@astro.uni-tuebingen.de}
           \and
           Astrophysique et Spectroscopie, Universit\'e de Mons -- UMONS, 7000 Mons, Belgium
           \and
           IPNAS, Universit\'e de Li\`ege, Sart Tilman, 4000 Li\`ege, Belgium
           \and
           Centro de Investigaciones Opticas (CIOp), CC 3, 1897, Gonnet, La Plata, Argentina}

\date{Received 12 March 2015; 8 April accepted 2015}

\abstract {For the spectral analysis of spectra of hot stars
with a high resolution and high signal-to-noise ratio (S/N),
           advanced non-local thermodynamic equilibrium (NLTE) 
           model atmospheres are mandatory. These are strongly
           dependent on the reliability of the atomic data that are used for their calculation.
          }
          {Reliable \ion{Xe}{vi} oscillator strengths 
           are used to identify Xe lines in the ultraviolet spectrum of 
           the DO-type white dwarf \re and
           to determine its photospheric Xe abundance.
          }
          {We publish newly calculated oscillator strengths that
are based on a recently measured \ion{Xe}{vi} laboratory line spectrum. These strengths were used
           to consider their radiative and collisional bound-bound transitions
           in detail in our NLTE stellar-atmosphere models
           to analyze \ion{Xe}{vi} lines exhibited in
           high-resolution and high S/N UV observations of \re.
          }
          {We identify three hitherto unknown \ion{Xe}{vi} lines in the ultraviolet spectrum of 
           \re and confirm the previously measured photospheric Xe abundance of this white dwarf 
           ($\log\, \mathrm{Xe} = -4.2 \pm 0.6$ by mass).
           }
          {Reliable measurements and calculations of atomic data are prerequisite for stellar-atmosphere modeling. 
           Observed \ion{Xe}{vi} line profiles in the ultraviolet spectrum of the white dwarf \re 
           were well reproduced with the newly calculated \ion{Xe}{vi} oscillator strengths. 
          }


\maketitle

\section{Introduction}
\label{sect:intro}

\citet{werneretal2012} identified lines of ten trans-iron elements in the 
Far Ultraviolet Spectroscopic Explorer (FUSE\footnote{\url{http://fuse.pha.jhu.edu}}) observations of \re. They 
were only able to determine the photospheric krypton and xenon abundances because 
reliable transition probabilities were solely available for these species.
New calculations of transition probabilities enabled us in the meantime to measure the abundances of
Zn, Ge, Ga, and Ba \citep[][respectively]{rauchetal2014zn,rauchetal2012ge,rauchetal2014ba,rauchetal2015ga}.

Recent measurements of \citet{gallardoetal2015} have doubled the
spectral information about \ion{Xe}{vi}. Since reliable spectral analyses
by means of non-local thermodynamic equilibrium (NLTE) model-atmosphere
calculations need reliable transition probabilities for the complete model atom
(and not only for the identified lines),
these new data allow improving the abundance determination of Xe.
Moreover, the many newly classified \ion{Xe}{vi} lines enable us to
assign hitherto unidentified lines in the observation.

We start with an introduction to our observations and atmosphere models (Sects.\,\ref{sect:obs} and \ref{sect:models}).
In Sect.\,\ref{sect:xetrans}, we briefly compare the new \ion{Xe}{vi} transition probabilities
to previous measurements. The line identification and abundance analysis is described in Sect.\,\ref{sect:lineid}.
We summarize the results in Sect.\,\ref{sect:results}.

\section{Observations}
\label{sect:obs}

We analyzed the same FUSE spectrum 
($910\,\mathrm{\AA} < \lambda <  1188\,\mathrm{\AA}$, resolving power $R = \lambda/\Delta\lambda \approx 20\,000$) 
that was previously described in detail by \citet{werneretal2012}.
In addition, we used our recently obtained (and later co-added) 
Hubble Space Telescope / Space Telescope Imaging Spectrograph 
(HST/STIS\footnote{\url{http://www.stsci.edu/instruments/stis}}) spectra
(ObsIds OC7N01010, OC7N01020, grating E140M, 
$1144\,\mathrm{\AA} < \lambda <  3073,\mathrm{\AA}$, $R \approx 45\,800$,
2014-08-14, total exposure time of 5494\,s).
The spectra are available via the Barbara A\@. Mikulski Archive for Space Telescopes 
(MAST\footnote{\url{http://archive.stsci.edu/}}).

\begin{figure*}
   \resizebox{\hsize}{!}{\includegraphics{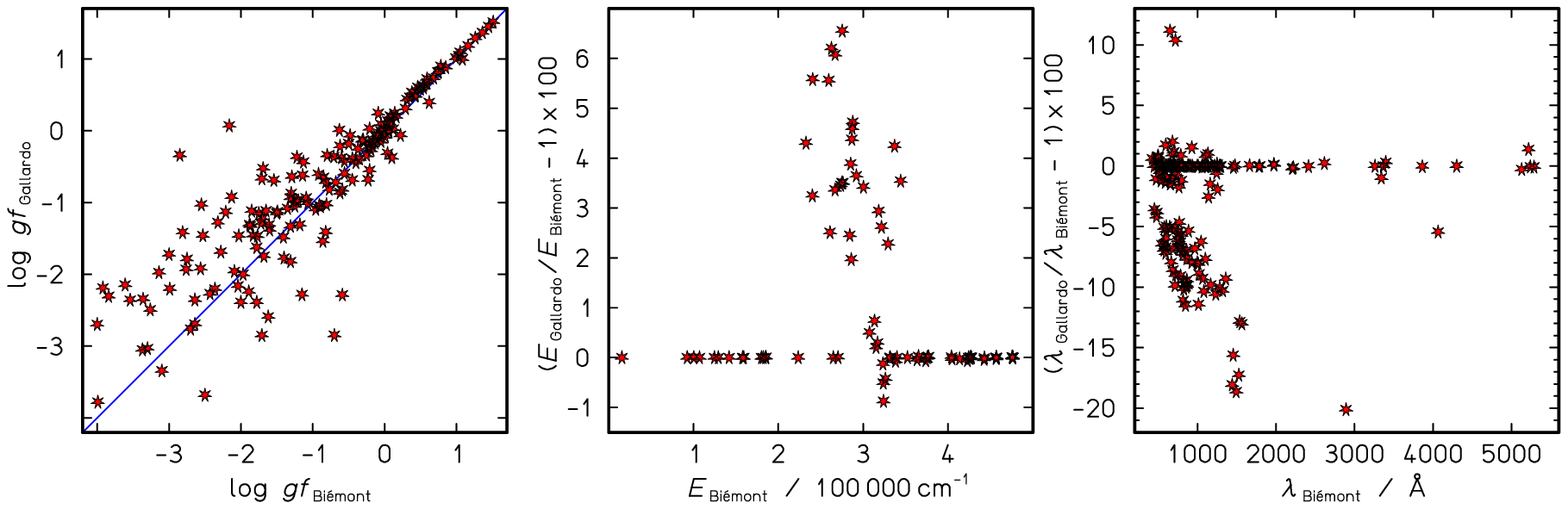}}
    \caption{Comparison of \ion{Xe}{vi} data 
             given by
              \citet[B+05]{biemontetal2005} and 
              \citet[G+15]{gallardoetal2015}.
             Left: 
             weighted oscillator strengths ($\log gf$) in G+15 as a function of the B+05 values,
             middle: 
             ratio of atomic level energies as a function of the B+05 values,
             right: 
             ratio of line wavelengths as a function of the B+05 values.
            }
   \label{fig:comp}
\end{figure*}

\section{Model atmospheres}
\label{sect:models}

Our model atmospheres are plane-parallel, chemically homogeneous,
and in hydrostatic and radiative equilibrium. They are calculated 
with the T\"ubingen NLTE Model Atmosphere Package
\citep[TMAP\footnote{\url{http://astro.uni-tuebingen.de/~TMAP}},][]{werneretal2003,tmap2012}.
Model atoms were constructed from the T\"ubingen Model Atom Database
\citep[TMAD\footnote{\url{http://astro.uni-tuebingen.de/~TMAD}},][]{rauchdeetjen2003}.
TMAD has been constructed as part of the T\"ubingen contribution to the German Astrophysical Virtual Observatory 
(GAVO\footnote{\url{http://www.g-vo.org}}).

\section{\ion{Xe}{vi} transition probabilities}
\label{sect:xetrans}

For their analysis of \re, \citet{werneretal2012} used \ion{Xe}{vi} data of \citet{reynaalmandosetal2001}, who
used the Hartree-Fock relativistic (HFR) approach to calculate weighted oscillator strengths ($gf$) of 104 
spectral lines belonging to the 5s5p$^2$, 5s$^2$5d, 5s$^2$6s, 5s5p5d, 5p$^3$, and 5s5p6s transitions array. 
In addition, $gf$-values of an experimental and theoretical study \citet{biemontetal2005} were used. 
These replaced $gf$-values of \citet{reynaalmandosetal2001} where available. They were calculated for 
$\Delta n = 0$ and $\Delta n = 1$ transitions connecting the 
5s$^2$nl [ np (n=5-8); nf (n=4-5); nh (n=6-8); nk (n=8) ] , 
5s5pnl (nl=5d,6s), 
5p$^3$, 
5s$^2$nl [ ns (n=6-8); nd (n=5-8); ng (n=5-6); ni (n=7-8) ], and 
5s5p$^2$ 
configurations. 
Core-polarization (CP) effects were included in their HFR approach. 
Good agreement was observed between theory and experiment.

\citet{gallardoetal2015} observed 243 lines (146 classified for the first time) of the \ion{Xe}{vi} spectrum within
$400\,\mathrm{\AA} \le \lambda \le 5500\,\mathrm{\AA}$. They calculated $gf$-values of
5s5p5d and
5s5p6s configurations.
\citet{biemontetal2005} encountered a problem in matching the 6d experimental level in their least-squares
calculations, which \citet[see this paper for further technical details]{gallardoetal2015} solved by adjusting 
configuration interaction (CI) integrals. They achieved standard deviations of 149\,cm$^{-1}$ 
in the odd parity energies
(35 configurations) 
and 154\,cm$^{-1}$ 
(34 configurations)
in the even parity.

We updated our \ion{Xe}{vi} model ion \citep{werneretal2012} with their experimental level energies (88 levels,
32 classified for the first time, 33 with revised energies)
and oscillator strengths (used in the formulae for the cross-sections of radiative and collisional bound-bound transitions). 
The new \ion{Xe}{vi} model ion is available via TMAD.

Figure \ref{fig:comp} shows the comparison
of the  \ion{Xe}{vi} $\log gf$ values, level energies, and line wavelengths of \citet{biemontetal2005} and the \citet{gallardoetal2015}. 
For the strongest lines, that is, $\log gf > 0.0,$ the agreement of the oscillator strengths is very good. The energy levels
deviate significantly in the $230\,000\,\mathrm{cm^{-1}} < E < 350\,000\,\mathrm{cm^{-1}}$ interval. This 
results in stronger differences of the lines' wavelengths 
in the ultraviolet wavelength range ($\lambda < 1700\,\mathrm{\AA}$).
The following comparison will verify the quality of the new data.

We start reproducing the lines shown by \citet{werneretal2012}.
Table\,\ref{tab:loggf} shows a comparison of the $\log gf$ values that they used with the new values.

\begin{figure}
   \resizebox{\hsize}{!}{\includegraphics{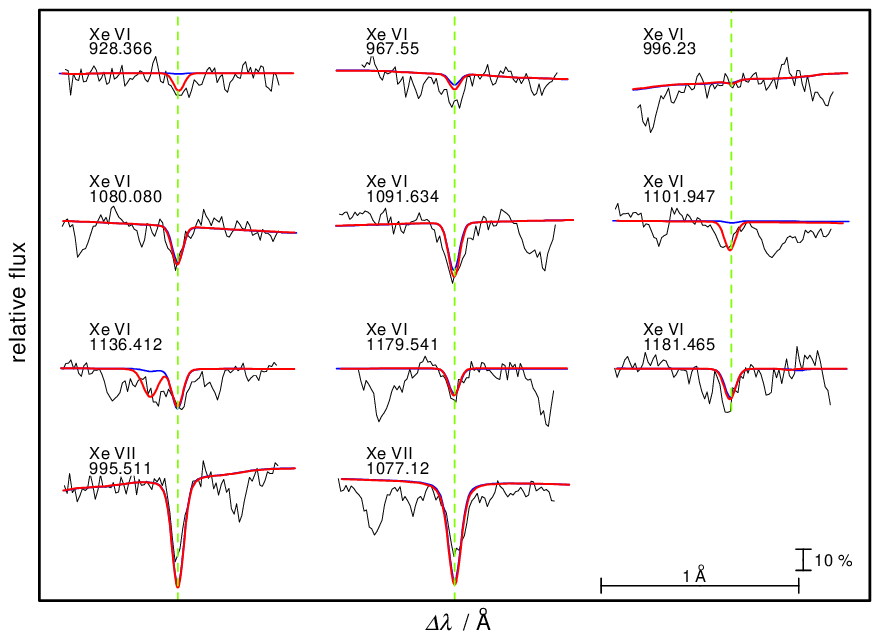}}
    \caption{\ion{Xe}{vi} and \ion{Xe}{vii} lines calculated from TMAP model atmospheres with
             \Teffw{70\,000}, \loggw{7.5}, and $\log\, \mathrm{Xe} = -4.2$ (mass fraction) compared with the FUSE
              observation. Thin (blue in the online version): 
              model of \citet[][\Jonw{N}{iv}{1136.27} was not included in the calculation]{werneretal2012}, thick (red) our model.
              The vertical bar indicates 10\,\% of the continuum flux level.
            }
   \label{fig:wernerlines}
\end{figure}

Figure\,\ref{fig:wernerlines} demonstrates how the new oscillator strengths influence the line-profile
fits. Because of their much higher $\log gf$ values, \Jonww{Xe}{vi}{928.366, 1101.947} now match the observation
significantly better, while the other lines are almost unchanged. The \ion{Xe}{vii} lines are too strong, 
most likely as a result of our insufficient \ion{Xe}{vii} model ion, for which only two oscillator strengths are known 
(namely of these two lines). In Sect.\,\ref{sect:lineid}, we search in the FUSE and HST/STIS
spectra for further \ion{Xe}{vi} lines to evaluate the oscillator strengths.

\begin{table}\centering
\caption{Comparison of the $\log gf$ values of \ion{Xe}{vi} lines 
         calculated by \citet{biemontetal2005} and used by \citet{werneretal2012} with
         those calculated by \citet{gallardoetal2015}.}
\label{tab:loggf}
\begin{tabular}{l@{\ --\ }lr@{.}lr@{.}lr@{.}l}
\hline
\hline
\noalign{\smallskip}
\multicolumn{4}{c}{~}                                                  & \multicolumn{4}{c}{$\log gf$}                               \\
\noalign{\smallskip}
\cline{5-8}
\multicolumn{8}{c}{~}                                                                                                 \vspace{-6.0mm}\\
\multicolumn{2}{c}{transition} & \multicolumn{2}{c}{$\lambda$\,/\,\AA} & \multicolumn{4}{c}{~}                        \vspace{-1.5mm}\\
\multicolumn{4}{c}{~}                                                   & \multicolumn{2}{c}{Bi{\'e}mont}   & \multicolumn{2}{c}{Gallardo}\\
\noalign{\smallskip}
\hline
\noalign{\smallskip}
$\mathrm{5s5p^2 }$ \hspace{0.1mm} $\mathrm{^2D_{3/2}          }$ &  $\mathrm{5p^3   }$ \hspace{3.2mm} $\mathrm{^2D_{3/2}^\mathrm{o}} $ &  928&37 & $-2$&$11$ & $-0$&$649$ \\
$\mathrm{5s5p^2 }$ \hspace{0.1mm} $\mathrm{^2D_{5/2}          }$ &  $\mathrm{5p^3   }$ \hspace{3.2mm} $\mathrm{^2D_{3/2}^\mathrm{o}} $ &  967&55 & $-1$&$14$ & $-0$&$937$ \\
$\mathrm{5s^2 5p}$ \hspace{0.1mm} $\mathrm{^2P_{1/2}^\mathrm{o}}$ &  $\mathrm{5s5p^2 }$ \hspace{0.1mm} $\mathrm{^4P_{3/2}          } $ &  996&23 & $-3$&$37$ & $-3$&$049$ \\
$\mathrm{5s^2 5p}$ \hspace{0.1mm} $\mathrm{^2P_{1/2}^\mathrm{o}}$ &  $\mathrm{5s5p^2 }$ \hspace{0.1mm} $\mathrm{^4P_{1/2}          } $ & 1080&08 & $-1$&$64$ & $-1$&$529$ \\
$\mathrm{5s^2 5p}$ \hspace{0.1mm} $\mathrm{^2P_{3/2}^\mathrm{o}}$ &  $\mathrm{5s5p^2 }$ \hspace{0.1mm} $\mathrm{^4P_{5/2}          } $ & 1091&63 & $-1$&$25$ & $-1$&$045$ \\
$\mathrm{5s5p^2 }$ \hspace{0.1mm} $\mathrm{^2P_{1/2}          }$ &  $\mathrm{5p^3   }$ \hspace{3.2mm} $\mathrm{^2D_{3/2}^\mathrm{o}} $ & 1101&95 & $-2$&$13$ & $-0$&$606$ \\
$\mathrm{5s^2 5d}$ \hspace{0.1mm} $\mathrm{^2D_{5/2}          }$ &  $\mathrm{5s^26p }$ \hspace{0.1mm} $\mathrm{^2P_{3/2}^\mathrm{o}} $ & 1136&41 & $+0$&$06$ & $-0$&$005$ \\
$\mathrm{5s^2 5p}$ \hspace{0.1mm} $\mathrm{^2P_{3/2}^\mathrm{o}}$ &  $\mathrm{5s5p^2 }$ \hspace{0.1mm} $\mathrm{^4P_{3/2}          } $ & 1179&54 & $-2$&$09$ & $-1$&$958$ \\
$\mathrm{5s^2 5d}$ \hspace{0.1mm} $\mathrm{^2D_{3/2}          }$ &  $\mathrm{5s^26p }$ \hspace{0.1mm} $\mathrm{^2P_{1/2}^\mathrm{o}} $ & 1181&46 & $-0$&$21$ & $-0$&$226$ \\
\hline
\end{tabular}
\end{table}

\section{Line identification and abundance determination}
\label{sect:lineid}

\begin{figure}
   \resizebox{\hsize}{!}{\includegraphics{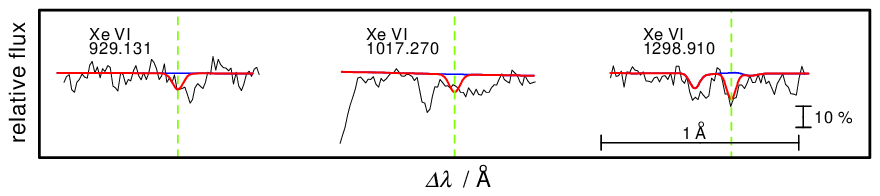}}
    \caption{\ion{Xe}{vi} lines calculated from the same models as in Fig.\,\ref{fig:wernerlines} compared with the FUSE and STIS 
              (right panel, \Jonw{Ni}{v}{1298.738} appears left of \Jonw{Xe}{vii}{1298.910})
              observations. Thin (blue in the online version): 
              model without Xe, thick (red) model including Xe.
            }
   \label{fig:rauchlines}
\end{figure}

We compared our best model spectra with our FUSE and STIS observations to identify additional \ion{Xe}{vi} lines.
We found only a few (Fig.\,\ref{fig:rauchlines}). \Jonww{Xe}{vi}{929.121, 1017,270} have 
apparently blends with so far unknown lines. \Jonw{Xe}{vi}{1298.910} matches the STIS observation well.
Since all observed Xe lines are reproduced well,  we confirm
the abundance
determination by \citet[][$\log\, \mathrm{Xe} = -4.2 \pm 0.6$]{werneretal2012}.
The two identified \Jonww{Xe}{vii}{995.511, 1077.12} lines 
\citep[oscillator strengths from][respectively]{kernahanetal1980,biemontelal2007}
are stronger than observed at this abundance (hence the 
relatively broad error range), but this may be an artifact due to the lack of reliable \ion{Xe}{vii}
oscillator strengths, cf\@. \citet[see Sect.\,\ref{sect:intro} above]{werneretal2012}.

\section{Results and conclusions}
\label{sect:results}

The recently presented oscillator strengths of \citet[][243 lines]{gallardoetal2015} complement and improve available
data of 
\citet[][104 lines]{reynaalmandosetal2001} and
\citet[][491 lines]{biemontetal2005}. A comparison of data for lines that
are listed in \citet{gallardoetal2015} and \citet{biemontetal2005}
showed that the $\log gf$ values agree well for the strongest lines. Deviations
in the line wavelengths are noted, but they do not affect the identified lines in
the UV spectra of \re and cannot be assessed. However, a comprehensive recalculation of
\ion{Xe}{vi} transition probabilities is desirable.

The oscillator strengths of \citet{gallardoetal2015} significantly
improve the agreement of the synthetic \ion{Xe}{vi} line profiles
with the observations in two of nine cases. The abundance determination of
\citet[][$\log\, \mathrm{Xe} = -4.2 \pm 0.6$]{werneretal2012} was confirmed.

\begin{figure}
   \resizebox{\hsize}{!}{\includegraphics{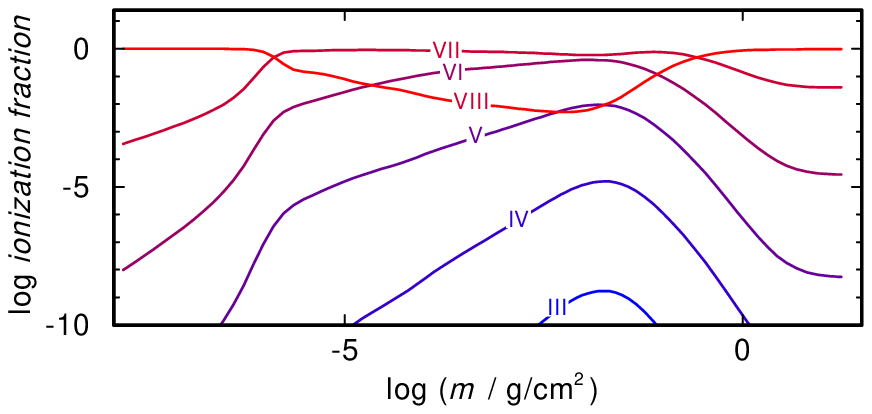}}
    \caption{\ion{Xe}{vi} ionization fractions in our best model for \re \citep{rauchetal2015ga}.
            }
   \label{fig:ionxevi}
\end{figure}

Figure\,\ref{fig:ionxevi} shows that \ion{Xe}{vii} is the dominant Xe ionization stage in the 
in the line-forming region ($-4 \la \log\,m \la 0.5$). 
$m$ is the mass column density, measured from the outer limit of the atmosphere.
Therefore, we expect to identify many more \ion{Xe}{vii} lines once reliable oscillator strengths 
for the complete \ion{Xe}{vii} model ion are available. This is mandatory to be able to reduce the uncertainty 
in the Xe abundance analysis.

\begin{acknowledgements}
TR and DH are supported by the German Aerospace Center (DLR, grants 05\,OR\,1402 and 50\,OR\,1501, respectively).
The GAVO project at T\"ubingen had been supported by the Federal Ministry of Education and
Research (BMBF, grants  05\,AC\,6\,VTB, 05\,AC\,11\,VTB).
Financial support from the Belgian FRS-FNRS is also acknowledged. 
PQ is research director of this organization.
Some of the data presented in this paper were obtained from the
Mikulski Archive for Space Telescopes (MAST). STScI is operated by the
Association of Universities for Research in Astronomy, Inc., under NASA
contract NAS5-26555. Support for MAST for non-HST data is provided by
the NASA Office of Space Science via grant NNX09AF08G and by other
grants and contracts. 
This research has made use of 
NASA's Astrophysics Data System and
the SIMBAD database, operated at CDS, Strasbourg, France.
\end{acknowledgements}

\bibliographystyle{aa}
\bibliography{26078}

\end{document}